
%
%
  \font\twelverm=cmr10 scaled 1200       \font\twelvei=cmmi10 scaled 1200
  \font\twelvesy=cmsy10 scaled 1200      \font\twelveex=cmex10 scaled 1200
  \font\twelvebf=cmbx10 scaled 1200      \font\twelvesl=cmsl10 scaled 1200
  \font\twelvett=cmtt10 scaled 1200      \font\twelveit=cmti10 scaled 1200
  \font\twelvemib=cmmib10 scaled 1200
  \font\elevenmib=cmmib10 scaled 1095
  \font\tenmib=cmmib10
  \font\eightmib=cmmib10 scaled 800
  
\font\elevenrm=cmr10 scaled 1095    \font\eleveni=cmmi10 scaled 1095
\font\elevensy=cmsy10 scaled 1095

%
%

\font\seventeeni=cmmi10 scaled \magstep3

\font\seventeensy=cmsy10 scaled \magstep3

\font\seventeenmib=cmmib10 scaled \magstep3

\newfam\cpfam%



\skewchar\eleveni='177   \skewchar\elevensy='60
\skewchar\elevenmib='177  \skewchar\seventeensy='60
\skewchar\seventeenmib='177
\skewchar\seventeeni='177

\newfam\mibfam%


  \skewchar\twelvei='177   \skewchar\twelvesy='60
  \skewchar\twelvemib='177
%
%
\def\twelvepoint{\normalbaselineskip=12.4pt
  \abovedisplayskip 12.4pt plus 3pt minus 9pt
  \belowdisplayskip 12.4pt plus 3pt minus 9pt
  \abovedisplayshortskip 0pt plus 3pt
  \belowdisplayshortskip 7.2pt plus 3pt minus 4pt
  \smallskipamount=3.6pt plus 1.2pt minus 1.2pt
  \medskipamount=7.2pt plus 2.4pt minus 2.4pt
  \bigskipamount=14.4pt plus 4.8pt minus 4.8pt
  \def\rm{\fam0\twelverm}          \def\it{\fam\itfam\twelveit}%
  \def\sl{\fam\slfam\twelvesl}     \def\bf{\fam\bffam\twelvebf}%
  \def\mit{\fam 1}                 \def\cal{\fam 2}%
  \def\tt{\twelvett}%
  \def\mib{\fam\mibfam\twelvemib}%

  \textfont0=\twelverm   \scriptfont0=\tenrm     \scriptscriptfont0=\sevenrm
  \textfont1=\twelvei    \scriptfont1=\teni      \scriptscriptfont1=\seveni
  \textfont2=\twelvesy   \scriptfont2=\tensy     \scriptscriptfont2=\sevensy
  \textfont3=\twelveex   \scriptfont3=\twelveex  \scriptscriptfont3=\twelveex
  \textfont\itfam=\twelveit
  \textfont\slfam=\twelvesl
  \textfont\bffam=\twelvebf
  \textfont\mibfam=\twelvemib       \scriptfont\mibfam=\tenmib
                                             \scriptscriptfont\mibfam=\eightmib

  \def\xrm{\textfont0=\twelverm\scriptfont0=\tenrm
      \scriptscriptfont0=\sevenrm\rm}
\normalbaselines\rm}


\mathchardef\alpha="710B
\mathchardef\beta="710C
\mathchardef\gamma="710D
\mathchardef\delta="710E
\mathchardef\epsilon="710F
\mathchardef\zeta="7110
\mathchardef\eta="7111
\mathchardef\theta="7112
\mathchardef\kappa="7114
\mathchardef\lambda="7115
\mathchardef\mu="7116
\mathchardef\nu="7117
\mathchardef\xi="7118
\mathchardef\pi="7119
\mathchardef\rho="711A
\mathchardef\sigma="711B
\mathchardef\tau="711C
\mathchardef\phi="711E
\mathchardef\chi="711F
\mathchardef\psi="7120
\mathchardef\omega="7121
\mathchardef\varepsilon="7122
\mathchardef\vartheta="7123
\mathchardef\varrho="7125
\mathchardef\varphi="7127

\def\physgreek{
\mathchardef\Gamma="7100
\mathchardef\Delta="7101
\mathchardef\Theta="7102
\mathchardef\Lambda="7103
\mathchardef\Xi="7104
\mathchardef\Pi="7105
\mathchardef\Sigma="7106
\mathchardef\Upsilon="7107
\mathchardef\Phi="7108
\mathchardef\Psi="7109
\mathchardef\Omega="710A}


\def\beginlinemode{\endmode
  \begingroup\parskip=0pt \obeylines\def\\{\par}\def\endmode{\par\endgroup}}
\def\beginparmode{\endmode
  \begingroup \def\endmode{\par\endgroup}}
\let\endmode=\par
{\obeylines\gdef\
{}}
\def\singlespace{\baselineskip=\normalbaselineskip}

\def\oneandahalfspace{\baselineskip=\normalbaselineskip
  \multiply\baselineskip by 3 \divide\baselineskip by 2}
\def\doublespace{\baselineskip=\normalbaselineskip \multiply\baselineskip by 2}

\nopagenumbers
\newcount\firstpageno
\firstpageno=2
\headline={\ifnum\pageno<\firstpageno{\hfil}\else{\hfil\elevenrm\folio\hfil}\fi}
\let\rawfootnote=\footnote             
\def\footnote#1#2{{\singlespace\parindent=0pt
\rawfootnote{#1}{#2}}}
\def\raggedcenter{\leftskip=4em plus 12em \rightskip=\leftskip
  \parindent=0pt \parfillskip=0pt \spaceskip=.3333em \xspaceskip=.5em
  \pretolerance=9999 \tolerance=9999
  \hyphenpenalty=9999 \exhyphenpenalty=9999 }
\def\dateline{\rightline{\ifcase\month\or
  January\or February\or March\or April\or May\or June\or
  July\or August\or September\or October\or November\or December\fi
  \space\number\year}}
\def\received{\vskip 3pt plus 0.2fill
 \centerline{\sl (Received\space\ifcase\month\or
  January\or February\or March\or April\or May\or June\or
  July\or August\or September\or October\or November\or December\fi
  \qquad, \number\year)}}


\hsize=6.5truein
\hoffset=0.0truein
\vsize=8.9truein
\voffset=0truein
\hfuzz=0.1pt
\vfuzz=0.1pt
\parskip=\medskipamount
\overfullrule=0pt      



\def\title                     
  {\null\vskip 3pt plus 0.1fill
   \beginlinemode \doublespace \raggedcenter \bf}

\def\author                    
  {\vskip 6pt plus 0.2fill \beginlinemode
   \singlespace \raggedcenter}

\def\affil        
  {\vskip 6pt plus 0.1fill \beginlinemode
   \oneandahalfspace \raggedcenter \it}

\def\abstract                  
  {\vskip 6pt plus 0.3fill \beginparmode
   \doublespace \narrower }

\def\summary                   
  {\vskip 3pt plus 0.3fill \beginparmode
   \doublespace \narrower SUMMARY: }

\def\pacs#1
  {\vskip 3pt plus 0.2fill PACS numbers: #1}

\def\endtitlepage              
  {\endpage                    
   \body}

\def\body                      
  {\beginparmode}              

\def\head#1{                   
  \filbreak\vskip 0.5truein    
  {\immediate\write16{#1}
   \raggedcenter \uppercase{#1}\par}
   \nobreak\vskip 0.25truein\nobreak}

%
%

%
\def\inlinerefs{
  \gdef\refto##1{ [##1]}                
\gdef\refis##1{\indent\hbox to 0pt{\hss##1.~}} 
\gdef\journal##1, ##2, ##3, 1##4##5##6{ 
    {\sl ##1~}{\bf ##2}, ##3 (1##4##5##6)}}    
\def\keywords#1
  {\vskip 3pt plus 0.2fill Keywords: #1}
\gdef\figis#1{\indent\hbox to 0pt{\hss#1.~}} 

\def\figurecaptions     
  {\head{Figure Captions}    
   \beginparmode
   \interlinepenalty=10000
   \frenchspacing \parindent=0pt \leftskip=1truecm
   \parskip=8pt plus 3pt \everypar{\hangindent=\parindent}}

%
%
\def\refto#1{$^{#1}$}          

\def\references       
  {\head{References}           
   \beginparmode
   \frenchspacing \parindent=0pt \leftskip=1truecm
   \interlinepenalty=10000
   \parskip=8pt plus 3pt \everypar{\hangindent=\parindent}}

\gdef\refis#1{\indent\hbox to 0pt{\hss#1.~}} 

\gdef\journal#1, #2, #3, 1#4#5#6{              
    {\sl #1~}{\bf #2}, #3 (1#4#5#6)}          

\def\refstylenp{               
  \gdef\refto##1{ [##1]}                               
  \gdef\refis##1{\indent\hbox to 0pt{\hss##1)~}}      
  \gdef\journal##1, ##2, ##3, ##4 {                    
     {\sl ##1~}{\bf ##2~}(##3) ##4 }}

\def\refstyleprnp{             
  \gdef\refto##1{ [##1]}                               
  \gdef\refis##1{\indent\hbox to 0pt{\hss##1)~}}      
  \gdef\journal##1, ##2, ##3, 1##4##5##6{              
    {\sl ##1~}{\bf ##2~}(1##4##5##6) ##3}}

\def\prb{\journal Phys. Rev. B, }

\def\prl{\journal Phys. Rev. Lett., }

\def\endreferences{\body}

%
%

\def\endfigurecaptions{\body}

\def\endpage                   
  {\vfill\eject}

\def\endpaper                  
  {\endmode\vfill\supereject}

\def\endit
  {\endpaper\end}


\def\ref#1{Ref.[#1]}                   
\def\Ref#1{Ref.[#1]}                   

\def\Equation#1{Equation [#1]}         
\def\Equations#1{Equations [#1]}       
\def\Eq#1{Eq. (#1)}                     
\def\eq#1{Eq. (#1)}                     
\def\Eqs#1{Eqs. (#1)}                   
\def\eqs#1{Eqs. (#1)}                   
\def\frac#1#2{{\textstyle{{\strut #1} \over{\strut #2}}}}

\def\sla{\raise.15ex\hbox{$/$}\kern-.57em}
\def\leaderfill{\leaders\hbox to 1em{\hss.\hss}\hfill}
\def\twiddle{\lower.9ex\rlap{$\kern-.1em\scriptstyle\sim$}}
\def\bigtwiddle{\lower1.ex\rlap{$\sim$}}
\def\gtwid{\mathrel{\raise.3ex\hbox{$>$\kern-.75em\lower1ex\hbox{$\sim$}}}}
\def\ltwid{\mathrel{\raise.3ex\hbox{$<$\kern-.75em\lower1ex\hbox{$\sim$}}}}
\def\square{\kern1pt\vbox{\hrule height 1.2pt\hbox{\vrule width 1.2pt\hskip 3pt
   \vbox{\vskip 6pt}\hskip 3pt\vrule width 0.6pt}\hrule height 0.6pt}\kern1pt}

%

%

%

%
\physgreek
%

\def\dsl{\raise.15ex\hbox{$/$}\kern-.57em\hbox{$\partial$}}
\def\nsl{\raise.15ex\hbox{$/$}\kern-.57em\hbox{$\nabla$}}
\def\gtwid{\,{\raise.3ex\hbox{$>$\kern-.75em\lower1ex\hbox{$\sim$}}}\,}
\def\ltwid{\,{\raise.3ex\hbox{$<$\kern-.75em\lower1ex\hbox{$\sim$}}}\,}
\def\undr{\raise.3ex\hbox{$\sim$\kern-.75em\lower1ex\hbox{$|\vec
x|\to\infty$}}}

\def\[{\left [}
\def\]{\right ]}
\def\({\left (}
\def\){\right )}







\def\and{a^{\phantom\dagger}}

%
\def\id{\raise.72ex\hbox{$-$}\kern-.85em\hbox{$d$}\,}

\catcode`@=11
\newcount\r@fcount \r@fcount=0
\newcount\r@fcurr
\immediate\newwrite\reffile
\newif\ifr@ffile\r@ffilefalse
\def\w@rnwrite#1{\ifr@ffile\immediate\write\reffile{#1}\fi\message{#1}}

\def\writer@f#1>>{}
\def\referencefile{
  \r@ffiletrue\immediate\openout\reffile=\jobname.ref%
  \def\writer@f##1>>{\ifr@ffile\immediate\write\reffile%
    {\noexpand\refis{##1} = \csname r@fnum##1\endcsname = %
     \expandafter\expandafter\expandafter\strip@t\expandafter%
     \meaning\csname r@ftext\csname r@fnum##1\endcsname\endcsname}\fi}%
  \def\strip@t##1>>{}}

\def\citeall#1{\xdef#1##1{#1{\noexpand\cite{##1}}}}
\def\cite#1{\each@rg\citer@nge{#1}}	

\def\each@rg#1#2{{\let\thecsname=#1\expandafter\first@rg#2,\end,}}
\def\first@rg#1,{\thecsname{#1}\apply@rg}	
\def\apply@rg#1,{\ifx\end#1\let\next=\relax
\else,\thecsname{#1}\let\next=\apply@rg\fi\next}

\def\citer@nge#1{\citedor@nge#1-\end-}	
\def\citer@ngeat#1\end-{#1}
\def\citedor@nge#1-#2-{\ifx\end#2\r@featspace#1 
  \else\citel@@p{#1}{#2}\citer@ngeat\fi}	
\def\citel@@p#1#2{\ifnum#1>#2{\errmessage{Reference range #1-#2\space is bad.}%
    \errhelp{If you cite a series of references by the notation M-N, then M and
    N must be integers, and N must be greater than or equal to M.}}\else%
 {\count0=#1\count1=#2\advance\count1
by1\relax\expandafter\r@fcite\the\count0,%
  \loop\advance\count0 by1\relax
    \ifnum\count0<\count1,\expandafter\r@fcite\the\count0,%
  \repeat}\fi}

\def\r@featspace#1#2 {\r@fcite#1#2,}	
\def\r@fcite#1,{\ifuncit@d{#1}
    \newr@f{#1}%
    \expandafter\gdef\csname r@ftext\number\r@fcount\endcsname%
                     {\message{Reference #1 to be supplied.}%
                      \writer@f#1>>#1 to be supplied.\par}%
 \fi%
 \csname r@fnum#1\endcsname}
\def\ifuncit@d#1{\expandafter\ifx\csname r@fnum#1\endcsname\relax}%
\def\newr@f#1{\global\advance\r@fcount by1%
    \expandafter\xdef\csname r@fnum#1\endcsname{\number\r@fcount}}

\let\r@fis=\refis			
\def\refis#1#2#3\par{\ifuncit@d{#1}
   \newr@f{#1}%
   \w@rnwrite{Reference #1=\number\r@fcount\space is not cited up to now.}\fi%
  \expandafter\gdef\csname r@ftext\csname r@fnum#1\endcsname\endcsname%
  {\writer@f#1>>#2#3\par}}

\def\ignoreuncited{
   \def\refis##1##2##3\par{\ifuncit@d{##1}%
     \else\expandafter\gdef\csname r@ftext\csname
r@fnum##1\endcsname\endcsname%
     {\writer@f##1>>##2##3\par}\fi}}

\def\r@ferr{\endreferences\errmessage{I was expecting to see
\noexpand\endreferences before now;  I have inserted it here.}}
\let\r@ferences=\references
\def\references{\r@ferences\def\endmode{\r@ferr\par\endgroup}}

\let\endr@ferences=\endreferences
\def\endreferences{\r@fcurr=0
  {\loop\ifnum\r@fcurr<\r@fcount
    \advance\r@fcurr by 1\relax\expandafter\r@fis\expandafter{\number\r@fcurr}%
    \csname r@ftext\number\r@fcurr\endcsname%
  \repeat}\gdef\r@ferr{}\endr@ferences}


\let\r@fend=\endpaper\gdef\endpaper{\ifr@ffile
\immediate\write16{Cross References written on []\jobname.REF.}\fi\r@fend}

\catcode`@=12

\citeall\refto		
\citeall\ref		%
\citeall\Ref		%

\catcode`@=11
\newcount\tagnumber\tagnumber=0

\immediate\newwrite\eqnfile
\newif\if@qnfile\@qnfilefalse
\def\write@qn#1{}
\def\writenew@qn#1{}
\def\w@rnwrite#1{\write@qn{#1}\message{#1}}
\def\@rrwrite#1{\write@qn{#1}\errmessage{#1}}

\def\taghead#1{\gdef\t@ghead{#1}\global\tagnumber=0}
\def\t@ghead{}

\expandafter\def\csname @qnnum-3\endcsname
  {{\t@ghead\advance\tagnumber by -3\relax\number\tagnumber}}
\expandafter\def\csname @qnnum-2\endcsname
  {{\t@ghead\advance\tagnumber by -2\relax\number\tagnumber}}
\expandafter\def\csname @qnnum-1\endcsname
  {{\t@ghead\advance\tagnumber by -1\relax\number\tagnumber}}
\expandafter\def\csname @qnnum0\endcsname
  {\t@ghead\number\tagnumber}
\expandafter\def\csname @qnnum+1\endcsname
  {{\t@ghead\advance\tagnumber by 1\relax\number\tagnumber}}
\expandafter\def\csname @qnnum+2\endcsname
  {{\t@ghead\advance\tagnumber by 2\relax\number\tagnumber}}
\expandafter\def\csname @qnnum+3\endcsname
  {{\t@ghead\advance\tagnumber by 3\relax\number\tagnumber}}

\def\equationfile{%
  \@qnfiletrue\immediate\openout\eqnfile=\jobname.eqn%
  \def\write@qn##1{\if@qnfile\immediate\write\eqnfile{##1}\fi}
  \def\writenew@qn##1{\if@qnfile\immediate\write\eqnfile
    {\noexpand\tag{##1} = (\t@ghead\number\tagnumber)}\fi}
}

\def\callall#1{\xdef#1##1{#1{\noexpand\call{##1}}}}
\def\call#1{\each@rg\callr@nge{#1}}

\def\each@rg#1#2{{\let\thecsname=#1\expandafter\first@rg#2,\end,}}
\def\first@rg#1,{\thecsname{#1}\apply@rg}
\def\apply@rg#1,{\ifx\end#1\let\next=\relax%
\else,\thecsname{#1}\let\next=\apply@rg\fi\next}

\def\callr@nge#1{\calldor@nge#1-\end-}
\def\callr@ngeat#1\end-{#1}
\def\calldor@nge#1-#2-{\ifx\end#2\@qneatspace#1 %
  \else\calll@@p{#1}{#2}\callr@ngeat\fi}
\def\calll@@p#1#2{\ifnum#1>#2{\@rrwrite{Equation range #1-#2\space is bad.}
\errhelp{If you call a series of equations by the notation M-N, then M and
N must be integers, and N must be greater than or equal to M.}}\else%
 {\count0=#1\count1=#2\advance\count1
 by1\relax\expandafter\@qncall\the\count0,%
  \loop\advance\count0 by1\relax%
    \ifnum\count0<\count1,\expandafter\@qncall\the\count0,%
  \repeat}\fi}

\def\@qneatspace#1#2 {\@qncall#1#2,}
\def\@qncall#1,{\ifunc@lled{#1}{\def\next{#1}\ifx\next\empty\else
  \w@rnwrite{Equation number \noexpand\(>>#1<<) has not been defined yet.}
  >>#1<<\fi}\else\csname @qnnum#1\endcsname\fi}

\let\eqnono=\eqno
\def\eqno(#1){\tag#1}
\def\tag#1$${\eqnono(\displayt@g#1 )$$}

\def\aligntag#1\endaligntag
  $${\gdef\tag##1\\{&(##1 )\cr}\eqalignno{#1\\}$$
  \gdef\tag##1$${\eqnono(\displayt@g##1 )$$}}

\def\eqalignno#1{\displ@y \tabskip\centering
  \halign to\displaywidth{\hfil$\displaystyle{##}$\tabskip\z@skip
    &$\displaystyle{{}##}$\hfil\tabskip\centering
    &\llap{$\displayt@gpar##$}\tabskip\z@skip\crcr
    #1\crcr}}

\def\displayt@gpar(#1){(\displayt@g#1 )}

\def\displayt@g#1 {\rm\ifunc@lled{#1}\global\advance\tagnumber by1
        {\def\next{#1}\ifx\next\empty\else\expandafter
        \xdef\csname @qnnum#1\endcsname{\t@ghead\number\tagnumber}\fi}%
  \writenew@qn{#1}\t@ghead\number\tagnumber\else
        {\edef\next{\t@ghead\number\tagnumber}%
        \expandafter\ifx\csname @qnnum#1\endcsname\next\else
        \w@rnwrite{Equation \noexpand\tag{#1} is a duplicate number.}\fi}%
  \csname @qnnum#1\endcsname\fi}

\def\ifunc@lled#1{\expandafter\ifx\csname @qnnum#1\endcsname\relax}

\let\@qnend=\end\gdef\end{\if@qnfile
\immediate\write16{Equation numbers written on []\jobname.EQN.}\fi\@qnend}

\catcode`@=12
\callall\Equation
\callall\Equations
\callall\Eq
\callall\eq
\callall\Eqs
\callall\eqs


\referencefile

\twelvepoint\doublespace

\title{Effect of Structure on the Electronic Density of States
of Doped Lanthanum Cuprate}

\author{M. R. Norman}
\affil
Materials Science Division
Argonne National Laboratory
Argonne, IL  60439

\author{G. J. McMullan}
\affil
Cavendish Laboratory
Madingley Road
Cambridge  CB3 OHE, UK

\author{D. L. Novikov and A. J. Freeman}
\affil
Science and Technology Center for Superconductivity and
Department of Physics and Astronomy
Northwestern University
Evanston, IL  60208

\abstract

We present a series of detailed band calculations on the various
structural phases of doped lanthanum cuprate:  HTT, LTO, and LTT.
The LTO distortion is shown to
have little effect on the electronic
density of states (DOS).  A fit to the
pressure dependence of the superconducting transition
temperature indicates that only 2.5\% of the DOS is
affected by the HTT$\rightarrow$ LTO transition.  The LTT distortion also
has little effect on the DOS for the experimental value of the
octahedral tilt angle.  Larger tilt angles, though, lead to a
dramatic change in the DOS.

\bigskip

\noindent PACS numbers:  71.25.Pi, 74.70.Vy

\endtitlepage

Doped lanthanum cuprate, $La_{2-x}M_xCuO_4$, where M is typically
Sr or Ba, is the prototype system for the class of
copper oxide materials known as
high temperature superconductors.  It exhibits a number
of structural phases, each of which has different superconducting
properties.  The HTT (high temperature body-centered tetragonal) phase
occurs for low temperatures only for x $>$ 0.2, where superconductivity
is suppressed.   At lower temperatures for x $<$ 0.2, one finds the LTO (low
temperature face-centered orthorhombic) phase, which is superconducting over a
range of x values.  Near x=0.125,
the LTT (low temperature primitive tetragonal) phase forms for the Ba system
with suppressed superconductivity.
A small dip in $T_c$ near x=0.115
is found in the Sr system, but no evidence for the LTT phase.
More information has
now been gathered by hydrostatic pressure experiments\refto{Yamada}.
For the range of x values where one has a superconducting LTO phase,
the HTT phase can be stabilized by pressure and is actually found to
have maximal $T_c$.  Near x=0.125 for the Ba system, pressure
destroys the LTT phase, yet superconductivity is still strongly
suppressed.

Understanding this series of puzzling results may help to unravel the
mystery behind high temperature superconductivity.  An obvious first
step in this direction is to understand the effect these
various structural distortions have on the electronic structure.  Of
course, many band structure calculations have been performed on these
systems in the past\refto{RMP}.  A recent calculation by Pickett
et al\refto{Pick} for the LTT phase revealed a strong suppression in the
density of states (DOS) near the Fermi energy ($E_F$), which they then
connected to the
suppressed superconductivity of this phase.  Because of this intriguing
result, and the various additional experimental phenomena mentioned
above, we decided to perform a series of band calculations for the
various phases, accurately calculate the DOS in the
vicinity of $E_F$, and attempt to connect these results to the
experimental observations.

We use the linearized muffin tin orbital method\refto{And},
including combined correction terms.  Three independent codes were employed
as checks, one of which contains non-spherical corrections to the potential
inside the muffin tins\refto{Meth} (all three codes gave comparable results).
The calculations presented in this paper are scalar relativistic, employ
the Hedin-Lundqvist form for the exchange-correlation potential,  and use
two empty spheres per formula unit (located at the (1/2,0,1/4) points in HTT
notation).  Doping was simulated by reducing the Z value of the La site.
Calculations were converged on a
90 k point mesh inside the irreducible wedge of the Brillouin zone.  For the
final
iteration, eigenvalues for 180 k points for
the HTT and LTO and 144 k points for the LTT phase were generated and the
results were fit using a Fourier series spline analysis.  The
spline fit was checked by plotting bands along various symmetry directions,
and then used to generate a DOS based on a
tetrahedral decomposition of the zone (around 1.6 million tetrahedra were
used).

The HTT calculation was done using the lattice parameters of
Cox et al\refto{Cox} for x=0.1 Ba at 295 K.  Four LTO calculations were
carried out, one which used the results\refto{Cox} for x=0.1 Ba at 91 K,
and three which used new results on the Sr system at 10 K for x values of
0.1, 0.15, and 0.2\refto{Hinks}.  Two LTT calculations were performed, one
which
used the results\refto{Cox} for x=0.1 Ba at 15 K, and another based on the
theoretical parameters of Pickett et al\refto{Pick} obtained by minimization
of the total energy\refto{thanks}.  The latter set of parameters has a tilt
angle of the copper
oxide octahedra about twice that of the former.

Before moving to the main discussion on the DOS, we first
comment on the matter of rigid band behavior.  This has been questioned
based on the fact that the Cu ion would prefer to be close to a $d^9$
configuration, and thus rigid band behavior may not be observed since the
states
at $E_F$ are a mixture of Cu d and O p states.  Our own doping
results, based on adjusting the Z value of the La nucleus, exhibit an
intriguing
behavior.  While we indeed found rigid band behavior in the DOS,
the charge density did not exhibit such behavior.  In particular,
the DOS has about 60\% Cu d character,
with the remainder mostly O p.  But comparing charges at 0\% doping and
10\% doping, only about 20\% of the change in charge came from the Cu d
orbitals, with
the remaining 80\% coming from the La site.  We should note that the charge
on the La site is almost all due to reanalysis of charge from the surrounding
O sites since the LMTO method uses overlapping spheres.  Our
speculation is that the change in potential on the La site due to the reduction
of the Z value causes the charge analysis on that site to change in order to
compenstate for
the charge loss due to doping, thus largely preserving the d count on the Cu
site.  This occurs, however, in such a way that rigid band behavior is
maintained in the DOS.

In Fig. 1, we show plots of the LTO and HTT DOS for the x=0.1
Ba calculation.  The HTT results were generated assuming LTO symmetry so as
to eliminate differences due to using different Brillouin zones.  As one can
see, there are virtually no differences in the curves.  This has been further
verified by plots of the Fermi surface which show no detectable differences
between HTT and LTO (the "gaps" seen in the LTO Fermi surface plots in the
literature\refto{RMP} are simply a zone-fold back effect and have nothing
to do with the orthorhombic distortion).

We show plots of the LTO DOS for the x=0.1, x=0.15, and x=0.2 Sr
calculations in Fig. 2.  Again, there are virtually no differences in the
curves, indicating again that the orthorhombic distortion has only a weak
effect on the DOS (we note that the orthorhombic distortion
increases as x decreases).

Plots of the LTT and HTT DOS for the x=0.1 Ba calculation are presented
in Fig. 3.  The HTT results were generated assuming LTT symmetry
so as to eliminate differences due to the differing Brillouin zones.  The zero
of energy was set at 12.5\% doping, where the LTT phase is seen experimentally.
Again, there are virtually no differences in the DOS.
This indicates that the suppression of $T_c$ for the LTT phase is probably not
connected with a density of states effect.

In Fig. 4, we show plots of our LTT calculation for x=0.1 Ba versus a
calculation done using the lattice parameters of the previous work of Pickett
et al\refto{Pick}.  We note that the octahedral tilt angle used in their
work is about a factor of two larger than what we used based on the Cox et al
parameters\refto{Cox}.  One can see that the van Hove peak is split
with the Pickett et al\refto{Pick} parameters (our DOS plot is very
similar to theirs).  The van Hove peak is also split with our
choice of parameters, but the effect is too small to be noticeable in the DOS.
This difference occurs because the splitting of
the van Hove peak depends quadratically on the tilt angle\refto{thanks}.  The
large splitting in the Pickett et al\refto{Pick} case gives a notch in the DOS
close to $E_F$ which led them to suspect that this might be
responsible for the suppressed superconductivity.  Because of this strong
dependence on
tilt angle, it is of some importance for experimentalists to attempt to
accurately determine the octahedral tilt angle for the LTT structure.

We conclude this part by remarking that the LTO and LTT structural
distortions have
little effect on the DOS,
though large differences are found for the LTT case
with increased octahedral tilt angle.  Moreover,
improved sampling of the zone by using more calculated
k points acts to sharpen the calculated van Hove singularity (thus, most
published plots of the DOS underestimates the height of this
peak).  We should also remark that the LMTO calculations place the van
Hove singularity at about 21-22\% doping, whereas FLAPW calculations
place this peak at about 17\% doping\refto{RMP}.  We have found that
LMTO calculations which do not include the combined correction terms
place the van Hove peak at the same doping as the FLAPW calculations,
indicating that the location of the peak is sensitive to details of the
electronic structure calculation.

We now attempt to connect some of these observations with experiment.  We
start with the LTO$\rightarrow$ HTT transition induced by
pressure\refto{Yamada}.
$T_c$ increases linearly with pressure, then saturates at this transition.
The pressure dependence of the structural transition can be estimated
from anomalies in the thermal expansion.  To describe this, we employ
a theory due to Bilbro and McMillan\refto{Bil}.  This theory assumes that
the superconducting pair potential is independent of pressure, and that
the pressure dependence comes from a competition between a DOS
change caused by the structural distortion and the formation of a
superconducting gap.  This involves solving two coupled mean field equations
involving the superconducting gap and the charge density wave gap (assumed
to only occur over part of the Fermi surface).  At a pressure
where the two transitions merge, the ratio of $dT_s/dp$ to $dT_c/dp$ (where
$T_s$ is the structural transition temperature, $T_c$ the superconducting
transition temperature, and $p$ the pressure) is equal
to $-(N-N_1)/N_1$ (where $N$ is the total DOS, and $N_1$ is
that part of the DOS removed by the structural distortion).  The data for
both x=0.17 and x=0.19 (where some information exists
for estimating the pressure dependence of the structural transition) give
values of $dT_c/dp$ $\simeq$ -5.75 K/kbar and $dT_c/dp$ $\simeq$ 0.15 K/kbar.
Thus,
$N_1/N$ $\simeq$ 0.025, i.e., only 2.5\% of the $N(E_F)$ value is affected
by the structural transition.  Such a small number is within the error bars
of the band calculations, and thus the data independently support our
conclusion that the LTO
distortion has a very weak effect on the DOS.  Moreover, this
theory would also predict that for pressures where the structural transition
is near (but larger than) $T_c$, one should see a saturation of the
orthorhombic distortion for $T < T_c$.  This effect should be observable
by neutron scattering experiments.

As for the LTO$\rightarrow$ LTT phase transition and the resultant
suppression of
superconductivity, our conclusion based on this work is that the density of
states does not play an important role.  This is consistent
with the pressure data, which show that even when the LTT transition is
gone, superconductivity is still suppressed.  Recent data\refto{msr} indicate
that magnetic ordering occurs for this concentration range, and thus is the
most likely reason for the $T_c$ suppression.  Given that band structure
calculations do not give rise to a magnetic transition for stochiomentric
$La_2CuO_4$, we do not expect to be able to describe this
magnetism.  As has been noted, the magnetism
may be due to a commensuration effect at x=1/8.  More theoretical work
is certainly needed to address this interesting effect.

In conclusion, we find little effect of the structural distortions of
doped $La_2CuO_4$ on the electronic density of states.  This result
is supported by some experimental data on both the HTT$\rightarrow$ LTO and
LTO$\rightarrow$ LTT transitions.

This work was supported by the National Science Foundation (DMR 91-20000)
through the Science and Technology Center for Superconductivity.
MRN was also supported by the U.S. Department of Energy, Office of Basic Energy
Sciences, under Contract No. W-31-109-ENG-38.  MRN would like to
acknowledge the hospitality of the Cavendish Laboratory, Cambridge
University,  where some of
this work was performed (with addtional support from the UK SERC and Trinity
College, Cambridge).  We thank Jim Jorgensen and Bernd Schuttler for calling
our attention to this problem and suggesting these calculations.  We also
acknowledge helpful conversations with Dale Koelling, and with
Mike Crawford and David Hinks concerning the experimental
data.  We are also indebted to Warren Pickett and Ron Cohen for many
correspondences concerning the LTT results.

\vfill\eject

\references

\refis{Yamada} N. Yamada and M. Ido, Physica C {\bf 203}, 240 (1992).

\refis{RMP} W.E. Pickett, Rev. Mod. Phys. {\bf 61}, 433 (1989).

\refis{Pick} W.E. Pickett, R.E. Cohen, and H. Krakauer, \prl 67, 228, 1991.

\refis{And} O.K. Andersen, \prb 12, 3060, 1975.

\refis{Meth} M. Methfessel, C.O. Rodriguez, and O.K. Andersen, \prb 40,
2009, 1989.

\refis{Cox} D.E. Cox, P. Zolliker, J.D. Axe, A.H.Moudden, A.R. Moodenbaugh,
and Y. Xu, Mat. Res. Soc. Symp. Proc. {\bf 156}, 141 (1989).

\refis{Hinks} David Hinks, private communication.

\refis{thanks} The authors thank Warren Pickett and Ron Cohen for their lattice
parameters and the observation about the quadratic dependence of the van Hove
splitting on tilt angle.

\refis{Bil} G. Bilbro and W.L. McMillan, \prb 14, 1887, 1976.

\refis{msr} I. Watanabe, K. Kawano, K. Kumagi, K. Nishiyama, and K. Nagamine,
J. Phys. Soc. Japan {\bf 61}, 3058 (1992).

\endreferences

\vfill\eject

\figurecaptions

\figis{1} Density of states (per formula unit) for the
HTT and LTO x=0.1 Ba cases.  The zero of energy is at
x=0.1

\figis{2} Density of states (per formula unit) for the
LTO x=0.1, 0.15, and 0.2 Sr cases.  The zero of
energy is at x=0.1

\figis{3} Density of states (per formula unit) for the
HTT and LTT x=0.1 Ba cases.  The zero of energy is at
x=0.125

\figis{4} Density of states (per formula unit) for the
LTT x=0.1 Ba case (Cox) and the LTT case with the
Pickett et al lattice parmaters (Pic).  The zero of
energy is at x=0.125

\endfigurecaptions

\vfill\eject

\endit